\documentclass[a4paper]{article}

\usepackage{latexsym}
\usepackage{float}
\floatstyle{ruled}
\newfloat{Algorithm}{thp}{lop}[section]

\newcommand{\qed}{\hfill$\Box$}

\usepackage{algorithm}
\usepackage{graphicx}
\usepackage{amssymb}
\usepackage[isolatin]{inputenc}
\usepackage[OT1]{fontenc}   
\usepackage{dsfont}
\usepackage{epsfig}
\usepackage{RR}

\newtheorem{definition}{Definition}
\newtheorem{theorem}{Theorem}
\newtheorem{lemma}{Lemma}

\newcommand{\BEGLIST}{\begin{list}{}{\partopsep -2pt \parsep -2pt \listparindent 0pt \labelwidth .5in}}
\newcommand{\ENDLIST}{\end{list}}

\RRNo{6458}

\RRdate{March 2008}

\RRtitle{Quiétude des commérages auto-stabilisants parmi des agents mobiles dans les graphes}
\RRetitle{Quiescence of Self-stabilizing Gossiping among Mobile Agents in Graphs}

\titlehead{Quiescence of Self-stabilizing Gossiping}

\RRauthor{Toshimitsu Masuzawa\thanks{Osaka University, Japan}
\and
Sébastien Tixeuil\thanks{Univ. Pierre \& Marie Curie - Paris 6, LIP6-CNRS \& INRIA, France}
}

\RRresume{Dans cet article, nous considérons le problème du commérage parmi des agents mobiles évoluant dans un graphe~: les agents se déplacent dans le graphe et doivent disséminer leur information initiale à tous les autres agents. Nous nous restreignons aux solutions auto-stabilisantes pour ce problème, où les agents peuvent démarrer depuis des positions arbitraires dans des états arbitraires. La propriété d'auto-stabilisation impose que certains des participants continuent de se déplacer indéfiniment, et mène au problème de maximiser le nombre d'agents autorisés à s'arrêter au bout d'un temps fini.

Cet article formalise le problème du commérage auto-stabilisant et introduit l'indice de quiétude (\emph{i.e.}, le nombre maximum d'agent autorisés à s'arrêter au bout d'un temps fini) des solutions auto-stabilisantes. Nous étudions l'évolution de l'indice de quiétude suivant les hypothèses liées à l'anonymat des agents, la synchronie, la bidirectionalité des liens, et la présence de tableaux blancs.}

\RRabstract{This paper considers gossiping among mobile agents in graphs: agents move on the graph and have to disseminate their initial information to every other agent.  We focus on self-stabilizing solutions for the gossip problem, where agents may start from arbitrary locations in arbitrary states.  Self-stabilization requires (some of the) participating agents to keep moving forever, hinting at maximizing the number of agents that could be allowed to stop moving eventually.

This paper formalizes the \emph{self-stabilizing agent gossip problem}, introduces the \emph{quiescence number} (\emph{i.e.}, the maximum number of eventually stopping agents) of self-stabilizing solutions and investigates the quiescence number with respect to several assumptions related to agent anonymity, synchrony, link duplex capacity, and whiteboard capacity.
}

\RRmotcle{Syst\`{e}mes distribu\'{e}s, Algorithme distribu\'{e}, Auto-stabilisation, Agents mobiles}
\RRkeyword{Distributed systems, Distributed algorithm, Self-stabilization, Mobile Agents}

\RRprojet{Grand large}
\RRtheme{\THNum}
\URFuturs

\begin{document}

\makeRR

\section{Introduction}

Distributed systems involving mobile entities called \emph{agents} or \emph{robots} recently attracted a widespread attention as they enable adaptive and flexible solutions to several problems. Intuitively, agents\footnote{Agents and robots can be used interchangeably in this paper.} are mobile entities operating in a network that is modeled by a graph; agents have limited computing capabilities and are able to move from a node to one of its neighbors.
The \emph{gossip problem} among mobile agents was introduced by Suzuki \emph{et al.}~\cite{SIOKM08,SIOKM07} as one of the most fundamental schemes supporting cooperation among mobile agents.
The problem requires each agent to disseminate the information it is initially given to all other agents.
Suzuki \emph{et al.}~\cite{SIOKM08,SIOKM07} investigated the problem of minimizing the number of agent moves for the gossip problem in \emph{fault-free} networks, and presented asymptotically optimal distributed solutions on several network topologies.

With the advent of large-scale networks that involve a total number of components in the order of the million, the fault (and attack) tolerance capabilities become at least as important as resource minimization.
In this paper, we consider the gossip problem in networks where both agents and nodes can be hit by unpredictable faults or attacks.
More precisely, we consider that arbitrary transient faults hit the system (both agents and nodes), and devise algorithmic solutions to recover from this catastrophic situation. The faults and attacks are \emph{transient} in the sense that there exists a point from which they don't appear any more. In practice, it is sufficient that the faults and attacks are sporadic enough for the network to provide useful services most of the time. 
Our solutions are based on the paradigm of \emph{self-stabilization}~\cite{D00b}, an elegant approach to forward recovery from transient faults and attacks as well as initializing large-scale systems. Informally, a self-stabilizing system is able to recover from any transient fault in finite time, without restricting the nature or the span of those faults.

\paragraph{Related works.}

Mobile (software) agents on graphs were studied in the context of self-stabilization \emph{e.g.} in \cite{BHS01c,DSW02c,G00c,HM01c}, but the implicit model is completely different from ours. In the aforementioned works, agents are software entities that are exchanged through messages between processes (that are located in the nodes of the network), and thus can be destroyed, duplicated, and created at will. The studied problems include stabilizing a network by means of a single non-stabilizing agent in~\cite{BHS01c,G00c}, regulating the number of superfluous agents in~\cite{DSW02c}, and ensuring regular traversals of $k$ agents in~\cite{HM01c}.

In this paper, we follow the model previously used in~\cite{BGT07c}, that studies necessary and sufficient condition for the problems of naming and electing agents in a network. The model assumes that the number of agents is fixed during any execution of the algorithm, but can start from any arbitrary location in the network and in any arbitrary initial state. Agents can communicate with other agents only if they are currently located on the same node, or make use of so-called \emph{whiteboards} - public memory variables located at each node. Of course, whiteboards may initially hold arbitrary contents due to a transient fault or attack. 

\paragraph{Our contribution.}

The contribution of this paper is twofold: 

\begin{enumerate}
\item We introduce the \emph{quiescence number} of self-stabilizing agent-based solutions to quantify communication efficiency after convergence. Self-stabilizing agent-based solutions inherently require (some of the) participating agents to \emph{keep moving} forever. This hints at maximizing the number of agents that could be allowed to stop moving after some point in every execution. The quiescence number denotes the maximum possible number of stopping agents\footnote{Minimizing communication after convergence in ordinary self-stabilizing solutions has been largely investigated with \emph{silent}~\cite{silent} protocols, and has been recently tackled, \emph{e.g.} in \cite{DDF07} for message passing systems} for a given particular problem.

\item We study the quiescence number of self-stabilizing $k$-gossiping (that denotes the gossiping among $k$ agents). The quiescence numbers we obtain are summarized in Table \ref{tab:results}, where "$-1$" represents impossibility of $0$-quiescence (that is, the problem is impossible to solve in a self-stabilizing way, even if agents are all allowed to move forever).
We consider the quiescence number under various assumptions about synchrony (synchronous/asynchronous), node whiteboards (\textbf{FW}/\textbf{CW}/\textbf{NW}), edge capacity (half-duplex/full-duplex) and anonymity of the agents. The details of the assumptions are presented in the next section.
\end{enumerate}

\begin{table}[tb]
\begin{center}
\caption{Quiescence numbers of the $k$-gossip problem}
\label{tab:results}
\tiny{
\begin{tabular}{|c|c|c|c|c|c|c|c|c|}
\hline
 & \multicolumn{4}{c|}{named agents} &
 \multicolumn{4}{c|}{anonymous agents} \\
\cline{2-9}
 & \multicolumn{2}{c|}{synchronous model} &
 \multicolumn{2}{c|}{asynchronous model} 
 & \multicolumn{2}{c|}{synchronous model} &
 \multicolumn{2}{c|}{asynchronous model} \\
\cline{2-9}
 whiteboards & half-duplex & full-duplex & half-duplex & full-duplex 
             & half-duplex & full-duplex & half-duplex & full-duplex \\
\hline
 \textbf{FW} & \multicolumn{2}{c|}{$k-1$ (Th. \ref{th:imp-k} \& \ref{th:prt-(k-1)SC})} & 
      \multicolumn{2}{c|}{$0$ (Th. \ref{th:qn-AF})} &
      \multicolumn{2}{c|}{$\ge 0$ (Th. \ref{th:qn-SFanonym})} & 
      \multicolumn{2}{c|}{$0$ (Th. \ref{th:qn-AFanonym})} \\
\hline
 \textbf{CW} & \multicolumn{2}{c|}{$k-1$ (Th. \ref{th:imp-k} \& \ref{th:prt-(k-1)SC})} & 
      0 (Th. \ref{th:qn-ACh}) & $-1$ (Th. \ref{th:qn-ACf}) &
      \multicolumn{2}{c|}{$-1$ (Th. \ref{th:qn-SCanonym})} & 
      \multicolumn{2}{c|}{$-1$ (Th. \ref{th:qn-SCanonym})} \\
\hline
 \textbf{NW} & \multicolumn{2}{c|}{$-1$ (Th. \ref{th:qn-N})} & 
      $-1$ (Th. \ref{th:qn-N}) & $-1$ (Th. \ref{th:qn-ACf}) &
      \multicolumn{2}{c|}{$-1$ (Th. \ref{th:qn-N})} & 
      $-1$ (Th. \ref{th:qn-N}) & $-1$ (Th. \ref{th:qn-ACf}) \\
\hline
\end{tabular}
}
\end{center}
\end{table}

\paragraph{Outline.}

In Section~\ref{sec:model}, we present the computing model with various
assumptions we consider in this paper. We also introduce the gossip 
problem and define the quiescence number of the gossip problem.
Section~\ref{sec:named} provides impossibility/possibility results 
in the model where each agent has a unique $id$.  
Section~\ref{sec:anonymous} briefly considers the quiescence numbers
in the model of anonymous agents.
Concluding remarks are presented in Section~\ref{sec:conclusion}.

\section{Preliminaries}
\label{sec:model}

\paragraph{Model.}

The network is modeled as a connected graph $G=(V,E)$, where $V$ is a set of nodes, and $E$ is a set of edges. We assume that nodes are \emph{anonymous}, that is, no node has a unique $id$ and all the nodes with the same degree are identical.
We also assume that nodes have local distinct labels for incident links, however no assumption is made about the labels. Each node also maintains a so-called \emph{whiteboard} that can store a fixed amount of information. 

\emph{Agents} (or \emph{robots}) are entities that move between neighboring nodes in the network. Each agent is modeled by a \emph{deterministic} state machine. An agent staying at a node may change its state, leave some information on 
the whiteboard of the node, and move to one of the node's neighbors based on the following information: \emph{(i)} the current state of the agent, \emph{(ii)} the current states of other agents located at the same node, \emph{(iii)} the local link labels of the current node (and possibly the label of the incoming link used by the agent to reach the node), and \emph{(iv)} the contents of the whiteboard at the node. In other words, the only way for two agents to communicate is by being hosted by the same node or by using node whiteboards.

In this paper, we consider several variants of the model, which fall into several categories: 
\begin{enumerate}
\item \textbf{Agent anonymity}: we consider two variants, named agents and anonymous agents. Each agent has a unique identifier in the \emph{named} agent model, and all agents are anonymous and identical in the \emph{anonymous} agent model.
\item \textbf{Synchrony}: we consider two variants, synchronous model and asynchronous model.
In the \emph{synchronous} model, all the agents are synchronized by rounds in the lock-step fashion.  Every agent executes its action at every round and can move to a neighboring node. When two (or more) agents are at the same node, all the agents execute their actions in one round but in sequence.
In the \emph{asynchronous} model, there is no bound on the number of moves that an agent can make between any two moves of another agent. However, we assume that each agent is eventually allowed to execute its action. When two (or more) agents are at the same node, they execute their actions sequentially. However, agents located at different nodes may execute their actions concurrently. 
\item \textbf{Link duplex capacity}: we consider two variants, full-duplex links and half-duplex links.  
A link is \emph{full-duplex} if two agents located at neighboring nodes can exchange their position at the same time, crossing the same link in opposite directions without meeting each other.  
A link is \emph{half-duplex} if only one direction can be used at a given time\footnote{If two agents at the different ends of a half-duplex link try to migrate along the link simultaneously, only one of them succeeds to migrate.}.

\item \textbf{Whiteboard capacity}: we consider three distinct hypothesis for information stored in the nodes' whiteboards. In the \textbf{NW} (No Whiteboard) model, no information can be stored in the whiteboard. In the \textbf{CW} (Control Whiteboard) model, only control information can be stored in the whiteboard. In the \textbf{FW} (Full Whiteboard) model, any information can be stored in the whiteboard (including gossip information, defined later in this section).
\end{enumerate}

Of course, there is a strict inclusion of the hypotheses, and a solution that requires only \emph{e.g.} the \textbf{NW} or the \textbf{CW} classes will work with the less restricted classes (\textbf{CW} and \textbf{FW}, and \textbf{FW}, respectively). Conversely if an impossibility result is shown for less restricted classes \emph{e.g.} \textbf{CW} and \textbf{FW}, it remains valid in the more restricted classes (\textbf{NW}, and \textbf{NW} and \textbf{CW}, respectively). 

The first set of hypotheses (or the agent anonymity) divides between Sections~\ref{sec:named} and \ref{sec:anonymous}. In each section, the remaining hypotheses (synchrony, link duplex capacity, and whiteboard capacity) are denoted by a tuple.  For example, ``$(Synch, \textbf{FW}, half)$-model'' denotes the synchronous model with \textbf{FW} whiteboards and half-duplex links. The wildcard ``$*$'' in the triplet denotes all possibilities for the category. For example, ``$(*, \textbf{FW}, half)$-model'' denotes both the $(Synch, \textbf{FW}, half)$-model and the $(Asynch, \textbf{FW}, half)$-model.

\paragraph{Gossip problem specification.}

We consider the \emph{gossip problem} among agents: agents are given some initial information (called \emph{gossip information}), and the goal of a protocol solving the problem is that each agent disseminates its gossip information to every other agent in the system. 
Each agent can transfer the gossip information to another agent by meeting it at a node or by leaving the gossip information in the whiteboard of a node. In the latter case, a \textbf{FW} whiteboard is required. The gossip information can be \emph{relayed} by other agents, that is, any agent that has already obtained the gossip information of another agent can transfer all collected gossip information to other agents and can store it in the whiteboard of a node.

In this paper, we consider \emph{self-stabilizing} solutions for the gossip problem. The solutions guarantee that every agent \emph{eventually} knows the gossip information of all the agents in the system even when the system is started from an \emph{arbitrary} configuration. So, agents may start from arbitrary locations in arbitrary states, and the nodes' whiteboards may initially contain arbitrary information. We assume that $k$ agents are present in the network at any time, yet $k$ is unknown to the agents. In the sequel, the $k$-\emph{gossip problem} denotes the problem of gossiping among $k$ agents.

\paragraph{Quiescence number.}

We introduce the \emph{quiescence} of solutions for the $k$-gossip problem to describe the fact that some agents, although executing local code, stop moving at some point of any execution.
 
\begin{definition}
A distributed algorithm for mobile agents is \emph{$l$-quiescent} (for some integer $l$) if any execution reaches a configuration after which $l$ (or more) agents remain still forever.
\end{definition}

\begin{definition}
The \emph{quiescence number} of a problem is the maximum integer $l$ such that a $l$-quiescent algorithm exists for the problem. For convenience, the quiescence number is considered to be $-1$ if there exists no $0$-quiescent algorithm (\emph{i.e.}, the problem is not solvable).
\end{definition}

Suzuki et al.~\cite{SIOKM08,SIOKM07} considered the \textbf{CW} and \textbf{FW} whiteboard models, and showed that the difference does not impact the move complexity of \emph{non-stabilizing} solutions for the gossip problem. In this paper, we clarify some differences among the \textbf{NW}, the \textbf{CW} and the \textbf{FW} whiteboard models with respect to the quiescence number of \emph{self-stabilizing} solutions for the gossip problem.

\section{Self-stabilizing $k$-gossiping among named agents}
\label{sec:named}

Our first result observes that the self-stabilization property of a $k$-gossiping protocol implies that at least one mobile agent must keep moving forever in the system.

\begin{theorem}
\label{th:imp-k}
There exists no $k$-quiescent self-stabilizing solution to the $k$-gossip problem in the $(*,*,*)$-model.
\end{theorem}

\noindent
\emph{Proof sketch.}
For the $k$-gossip problem to be self-stabilizing, there must not exist any \emph{terminal} configuration of agents, \emph{i.e.} a configuration from which all the agents never move thereafter. Assume the contrary, \emph{i.e.} there are $k$ agents (with $k\geq 2$) in a $n$-sized network (with $n>k$) such that agent and nodes are in a configuration such that all agents do not move thereafter. Then we construct a $2n$-sized network by mirroring the first network and joining the two by a node with no agent on it (there exists such a node since $n>k$). By mirroring the states and agents as well, there exists a second set of agents in the second part of the network that will never move thereafter. Since agents may only communicate by meeting other agents at the same node or by using whiteboards, the new $k$ agents from the second group are never able to communicate with any agent from the initial $k$-sized group. As agents do not have the knowledge of the actual number of agents in the system, none of them is able to distinguish between the two systems, hence the result.

The above discussion is valid independently of the assumptions concerning the synchrony, the link duplex capacity, or the whiteboard capacity.
\qed
\vspace{10pt}

Notice that Theorem \ref{th:imp-k} does not hold if agents know the number $k$ of existing agents. With the assumption of known $k$, it could be possible for agents to stop moving when $k$ agents are located at a same node, \emph{i.e.}, $k$-quiescence may be attainable if the \emph{rendez-vous} of $k$ agents is possible.
We now show that in the asynchronous case, no self-stabilizing algorithm can ensure that at least one agent does not move forever.
 
\begin{theorem}
\label{th:imp-lA}
There exists no $l$-quiescent self-stabilizing solution, for any $l\ (1 \le l \le k-1)$, to the $k$-gossip problem in $(Asynch, *, *)$-model.
\end{theorem}

\noindent
\emph{Proof sketch.}
Assume there exists a $1$-quiescent self-stabilizing solution, that is, in any execution there exists an agent that does not move after a certain configuration. The agent is only aware of the states of agents at the same node and the contents of the whiteboard at the node, and this information is sufficient to make the agent quiescent. Now consider that the graph is regular (\emph{i.e.} all nodes have the same degree) and non trivial. So, we can construct a configuration in which every agent is quiescent. Since we consider asynchronous systems, a quiescent agent cannot start moving unless another agent reaches the node. As a result, the agents never meet with each other and the gossiping cannot be achieved.
\qed
\vspace{10pt}

While Theorem \ref{th:imp-lA} precludes $l$-quiescence in \emph{asynchronous} models for any $l\ (1 \le l \le k-1)$, the impossibility result does not hold for \emph{synchronous} systems. 
Actually, in synchronous arbitrary networks, we present in Algorithms \ref{alg:(k-1)1}, \ref{alg:(k-1)2}, 
\ref{alg:(k-1)3}, and \ref{alg:(k-1)4} a positive result: a $(k-1)$-quiescent self-stabilizing solution to the $k$-gossip problem with \textbf{CW} whiteboards. 
The algorithm is based on the observation that gossiping can easily be achieved when a single agent repeatedly traverses the network: the agent alternates indefinitely a traversal to collect information and a traversal to distribute information. In our scheme, each agent may move according to a depth-first-traversal (DFT) in the network, and eventually an agent with minimal identifier (among all agents) keeps traversing forever, while other agents eventually stop. Since the network is synchronous, a stopped agent at node $u$ waits for the traversal of the minimal identifier agent a bounded period of time, then starts moving if no such agent visits $u$ within the bound. 

Each node $v$ has variables $InLink_v$ and $OutLink_v$ in its whiteboard to store information about the DFT of each agent $i$. We assume for simplicity that $v$ locally labels each incident link with an integer $a\ (0 \le a \le \Delta_v -1)$ where $\Delta_v$ is degree of $v$, and $v[a]$ denotes the neighbor of $v$ connected by the link labeled $a$. Variables $InLink_v$ and $OutLink_v$ have the following properties:

\begin{itemize}
\item A tuple $(i, a)\ (0 \le a \le \Delta_v -1)$ in variable $InLink_v$ of node $v$ implies that agent $i$ visited $v$ first from $v[a]$ (\emph{i.e}, $v[a]$ is the parent of $v$ in the depth-first-tree). A tuple $(i, \bot)$ in $InLink_v$ implies that $i$ did not visit $v$ yet, or that $i$ completed the DFT part starting from $v$ (and returned to the parent of $v$ in the depth-first-tree). For the starting node of the DFT, $(i, \bot)$ is always stored in $InLink_v$. We assume that only a single tuple of each agent $i$ can be stored in $InLink_v$ (this can be enforced having $InLink_v$ implemented through an associative memory) and we consider that the absence of any tuple involving $i$ denotes that $(i, \bot)$ is actually present.
\item A tuple $(i, a)\ (0 \le a \le \Delta_v -1)$ in variable $OutLink_v$ of node $v$ implies that agent $i$ left $v$ for $v[a]$ but did not return from $v[a]$ (\emph{i.e.}, $i$ is in the DFT starting from $v[a]$). A tuple $(i, \bot)$ in $OutLink_v$ implies that $i$ did not visit $v$ yet, or that $i$ completed the DFT part starting from $v$ (and returned to the parent of $v$ in the depth-first-tree). We assume the same additional constraints as for $InLink_v$.
\end{itemize}

In a legitimate configuration, tuples related to agent $i$ in $InLink_v$ and $OutLink_v$ of all nodes induce a path from the starting node to the currently visited node. However, in an arbitrary initial configuration, $InLink_v$ and $OutLink_v$ may contain arbitrary tuples for agent $i$ (several incomplete paths, cycles, no starting node, etc.).
We circumvent this problem by having each agent executing DFTs repeatedly. In order to distinguish the current DFT from the previous one, each agent $i$ maintains a boolean flag $t\_bit_i$ that is flipped when a new DFT is initiated.
Each node $v$ also maintains a variable $T\_table_v$ to store $t\_bit_i$ from the last visit of agent $i$ in the form of a tuple $(i, bit)$. For simplicity, we consider that $(i, true)$ is in $T\_table_v$ if no tuple of $i$ is contained in $T\_table_v$.

We now describe the mechanism to stop the remaining $k-1$ agents. We assume that each node $v$ maintains variables $MinID_v$, $WaitT_v$, and $Waiting_v$ in its whiteboard. The minimum $id$ among all agents having visited $v$ is stored in $MinID_v$, and the (computed) time required to complete a DFT is stored in $WaitT_v$. The completion time of a DFT is measured by the count-up timer $Timer_v$ of $v$ as follows. Agent $p$ with the minimum $id$ repeatedly makes DFTs. When visiting $v$ for the first time at each DFT, $p$ sets the count-up timer of $v$ to $WaitT_v$ and resets the timer. Eventually, $p$ completes each DFT in $2m$ rounds, where $m$ is the number of edges in the network, and $WaitT_v=2m$ remains true thereafter. When visiting $v$, an agent $p'$ finds a smaller $id$ in $MinID_v$ and stays at $v$ until the timer value of $v$ reaches $WaitT_v$. Since $p$ eventually completes each DFT in $2m$ rounds, each agent other than $p$ eventually remains at a node $v$ ($v$'s timer is reset regularly enough to never expire).

\begin{Algorithm}[tb]
{\small
\begin{tabbing}
xxx \= xxx \= xxx \= xxx \= xxx \= \kill
{\tt constants of agent $i$} \\
\> $i$: $id$ of $i$;  \\
{\tt constants of node $v$} \\
\> $deg_v$: degree of $v$;  \\
{\tt local variables of agent $i$} \\
\> $t\_bit_i$: {\tt bool;} \\
\> \> // an alternating bit to distinguish the current traversal from
         the previous one    \\
{\tt local variables of node $v$} \\
\> $T\_table_v:$ {\tt set of tuples $(id, t\_bit)$;} \\
\> \> // $(id, t\_bit)$ implies the latest visit of agent $id$ was done 
         with the alternating bit $t\_bit$    \\
\> $InLink_v:$ {\tt set of tuples $(id, port)$;} \\
\> \> // $(id, port)$ implies agent $id$ first came from $v[port]$ 
         in the current traversal    \\
\> \> // For each $id$, only the tuple updated last is stored \\
\> \> // $(id, \bot)$ is stored if $v$ is the initial node of the traversal \\
\> \> // $(id, \bot)$ is considered to be stored if no $(id, *)$ is present \\
\> $OutLink_v:$ {\tt set of tuples $(id, port)$;} \\
\> \> // $(id, port)$ implies agent $id$ went out from $v$ to $v[port]$ 
         last time it visited $v$ \\
\> \> // For each $id$, only the tuple updated last is stored \\
\> $MinID_v:$ {\tt agent id;} \\
\> \> // the minimum $id$ of the agents that have visited $v$ \\
\> $WaitT_v:$ {\tt int;} \\
\> \> // The amount of time agents with the non-minimum $id$ should wait \\
\> $Waiting_v:$ {\tt set of agents;} \\
\> \> // The set of agents waiting for timeout at $v$ \\
{\tt timers of node $v$} \\
\> $Timer_v$: {\tt count-up timer;} \\
\> \> The timer value is automatically increased by one at every round\\
{\tt functions on the local timer of node $v$} \\
\> $reset(Timer_v):$ Reset the timer value to 0 \\
\> $read(Timer_v):$ Return the timer value 
\end{tabbing}
}
\caption{Protocol (Part 1: constants, variables and timers)}
\label{alg:(k-1)1}
\end{Algorithm}

\begin{Algorithm}[htp]
{\small
\begin{tabbing}
xxx \= xxx \= xxx \= xxx \= xxx \= xxx \= \kill
{\tt Behavior of node $v$ at each round} \\
\> {\tt for each arriving agent $i$ do}\\
\> \> $visit_v(i)$; \\
\> $timeout\_check\_and\_execute_v$; 
\end{tabbing}
}
%
\caption{Protocol (Part 2: Main behavior)}
\label{alg:(k-1)2}
\end{Algorithm}

\begin{Algorithm}[htbp]
{\small
\begin{tabbing}
xxx \= xxx \= xxx \= xxx \= xxx \= xxx \= \kill
{\tt function} $visit_v(i)$; \\
\> // Executed when agent $i$ visits node $v$ from $v[a]$ 
      ($a$ can be any label initially \\
\> {\tt if} $((i, t\_bit_i) \not \in T\_table_v)$ \{ 
     // first visit of $i$ at $v$ in the current traversal \\
\> \> {\tt add $(i, t\_bit_i)$ to $T\_table_v$;} \\
\> \> {\tt add $(i, a)$ to $InLink_v$;} \\
\> \> {\tt if} $(i \le MinID_v)$ \{ \\
\> \> \> $MinID_v = i$; \\
\> \> \> $WaitT_v = read(Timer_v)$; \\
\> \> \> $reset(Timer_v)$; 
           // Timer is reset to start measuring the traversal time \\
\> \> \> {\tt if} $(deg_v \ge 2)$ \{ \\
\> \> \> \> {\tt add $(i, next_v(a))$ to $OutLink_v$;} 
              // $next_v(a) = (a+1)$ \textbf{mod} $deg_v$\\
\> \> \> \> {\tt migrate to} $v[next_v(a)]$; \\
\> \> \> \}  \\
\> \> \> {\tt else} \{ // $deg_v = 1$ then backtrack to $v[a]$ \\
\> \> \> \> {\tt add $(i, \bot)$ to $InLink_v$;} \\
\> \> \> \> {\tt migrate to} $v[a]$; \\
\> \> \> \} \\
\> \> \} \\
\> \> {\tt else} // $i > MinID_v$ \\
\> \> \> {\tt add $i$ to $Waiting_v$;} \\
\> \}  \\
\> {\tt else if} $((i, a) \not \in OutLink_v)$ \\
\> \> // $i$ previously visited $v$ in the current traversal,
         $i$ backtracks to $v[a]$. \\
\> \> \> {\tt migrate to} $v[a]$; \\
\> // The followings are the cases when $i$ backtracks to $v$ from $v[a]$. \\
\> {\tt else if $((next_v(a)==0)$ and $((i, \bot) \in InLink_v))$ \{} \\
\> \> // $v$ is the initial node of $i$'s traversal and 
         $i$ completes the current traversal \\
\> \> {\tt if} $(i \le MinID_v)$ \{ \\
\> \> \> $MinID_v = i$; $WaitT_v = read(Timer_v)$; $reset(Timer_v)$; \\
\> \> \> // Initiate a new traversal \\
\> \> \> $t\_bit_i = \neg t\_bit_i$; \\
\> \> \> {\tt add $(i, t\_bit_i)$ to $T\_table_v$;} \\
\> \> \> {\tt add $(i, 0)$ to $OutLink_v$;} \\
\> \> \> {\tt migrate to} $v[0]$; \\
\> \> \} \\
\> \> {\tt else} // $i > MinID_v$ \\
\> \> \> {\tt add $i$ to $Waiting_v$;} \\
\> \} \\
\> {\tt else if} $((i, next_v(a)) \in InLink_v)$ \{\\
\> \> // $v$ is not the initial node of $i$'s traversal,
         $i$ completes the current traversal from $v$ \\
\> \> {\tt add $(i, \bot)$ to $InLink_v$;} \\
\> \> {\tt add $(i, \bot)$ to $OutLink_v$;} \\
\> \> {\tt migrate to} $v[next_v(a)]$; // $i$ backtracks \\
\> \}  \\
\> {\tt else} \{ // $i$ did not complete the current traverse from $v$ \\
\> \> {\tt add $(i, next_v(a))$ to $OutLink_v$;} \\
\> \> {\tt migrate to} $v[next_v(a)]$; \\
\> \}  
\end{tabbing}
}
\caption{Protocol (Part 3: Behavior when agent $i$ arrives at node $v$ 
         from $v[a]$)}
\label{alg:(k-1)3}
\end{Algorithm}

\begin{Algorithm}[htbp]
{\small
\begin{tabbing}
xxx \= xxx \= xxx \= xxx \= xxx \= xxx \= xxx \= xxx \= \kill
{\tt function $timeout\_check\_and\_execute_v$;} \\
\> {\tt if} $read(Timer_v) \ge WaitT_v$; \{ // Timeout occurs \\
\> \> $MinID_v = \min \{j\ |\ j \in Waiting_v \}$; \\
\> \> {\tt Let} $i$ {\tt be such that} $MinID_v = i$; \\
\> \> $Waiting_v = Waiting_v - \{i\}$; \\
\> \> $reset(Timer_v)$; 
        // Timer is reset to start measuring the traversal time \\
\> \> {\tt if} $((i, a) \in InLink_v$ {\tt for some} 
      $a\ (0 \le a \le deg_v -1))$ \{ \\
\> \> \> // $v$ is not the initial node of $i$'s traversal \\
\> \> \> {\tt Let} $a$ {\tt be such that} $(i, a) \in InLink_v$; \\
\> \> \> {\tt if} $(deg_v \ge 2)$ \{ \\
\> \> \> \> {\tt add $(i, next_v(a))$ to $OutLink_v$;} \\
\> \> \> \> $i$ {\tt migrates to} $v[next_v(a)]$; \\
\> \> \> \}  \\
\> \> \> {\tt else} \{ // $deg_v = 1$ then backtrack to $v[a]$ \\
\> \> \> \> {\tt add} $(i, \bot)$ {\tt to} $InLink_v$; \\
\> \> \> \> $i$ {\tt migrates to} $v[a]$; \\
\> \> \> \}  \\
\> \> {\tt else} \{ // $v$ is the initial node of $i$'s traversal \\
\> \> \> // Initiate a new traversal \\
\> \> \> $t\_bit_i = \neg t\_bit_i$; \\
\> \> \> {\tt add $(i, t\_bit_i)$ to $T\_table_v$;} \\
\> \> \> {\tt add $(i, 0)$ to $OutLink_v$;} \\
\> \> \> {\tt migrate to $v[0]$;} \\
\> \> \} \\
\> \} 
\end{tabbing}
}
\caption{Protocol (Part 4: Behavior when Timeout occurs)}
\label{alg:(k-1)4}
\end{Algorithm}

\begin{lemma}
\label{lem:minID}
Starting from any initial configuration, in every execution of Algorithms~\ref{alg:(k-1)1}, \ref{alg:(k-1)2}, \ref{alg:(k-1)3}, and \ref{alg:(k-1)4}, eventually the agent with the minimum identifier repeatedly depth-first-traverses the network.
\end{lemma}

\noindent
\emph{Proof sketch.}
Let $p$ be the agent with the minimum $id$ (among all the agents in the system). When $p$ visits node $v$, if $p \le MinID_v$ then $MinID_v = p$ is executed. Otherwise (\emph{i.e.} when $MinID_v$ stores an identifier that is not the $id$ of any existing agent), $p$ suspends its DFT and waits for timeout at $v$ ($p$ is appended into $Waiting_v$). Since no agent with the \emph{fake} $id$ exists in the network, $read(Timer_v) \ge WaitT_v$ eventually holds (in function $timeout\_check\_and\_execute_v$). When this is the case, $MinID_v = \min \{j\ |\ j \in Waiting_v \}(=p)$ is executed and $p$ resumes the suspended DFT. Once $MinID_v$ is changed to $p$, $MinID_v$ never stores an $id$ smaller than $p$ again.

Now consider a DFT initiated by agent $p$ with $t\_bit_p=b\ (b\in\{true, false\})$.  In a legitimate configuration, $p$ initiates a DFT from a node $v$ satisfying $(p, \bot) \in InLink_v$. However, in the initial configuration, $(p, a) \in InLink_v$ may hold for some $a\ (0 \le a \le \Delta_v -1)$ where $v$ is the node $p$ is initially located at. We first show that $p$ eventually terminates the DFT starting from such an initial configuration and initiates a new DFT with $t\_bit_p=\neg b$. When $p$ with $t\_bit_p=b$ visits a node $u$ in a forward move, $p$ changes its tuple in $T\_table_u$ to $(p, b)$ if $(p, b) \not \in T\_table_u$. Otherwise, $p$ backtracks. Since $(p, b)$ in $T\_table_u$ never changes to $(p, \neg b)$ as long as $p$ continues the DFT with $t\_bit_p=b$, $p$ can make at most $m$ forward moves in the DFT.
On the other hand, agent $p$ backtracks from $u$ to $u[a]$ only when $(p,a) \in InLink_u$ holds. When backtracking from $u$ to $u[a]$, $p$ changes its tuple in $InLink_u$ to $(p,\bot)$. Thus, $p$ can make at most $n$ backtracking moves in the DFT. Consequently, $p$ eventually terminates the DFT even when it starts the DFT from a node $v$ with $(p, \bot) \not \in InLink_v$.

Now consider a DFT initiated by agent $i$ with $t\_bit_i=b$ at node $v$ with $(i, \bot) \in InLink_v$.
Let $G'=(V', E')$ be a connected component containing $v$ of $G^{\neg b}=(V^{\neg b}, E^{\neg b})$ where $V^{\neg b} = \{u \in V\ |\ (i, {\neg b}) \in T\_table_u$ when $i$ initiates the $DFT\}$ and $E^{\neg b} = (V^{\neg b} \times V^{\neg b}) \cap E$.  Since the algorithm can be viewed as a distributed version of a sequential DFT, it means $i$ makes a DFT in $G'$ and its outgoing edges (if they exist). When the DFT completes, the tuple of $i$ stored in $T\_table_u$ changes to $(i, b)$ at each $u$ in $V'$, while the tuple of $i$ stored in $T\_table_w$ remains unchanged at $w\ (\not \in V')$ during the DFT. Thus, if $G'$ is not the whole network, the connected component $G''$ (similarly defined as $G'$ for the next DFT with $t\_bit_i=\neg b$) contains at least one more node than $G'$. Since the network is finite, eventually $i$ makes DFTs repeatedly over the whole network.
\qed
\vspace{10pt}

\begin{theorem}
\label{th:prt-(k-1)SC}
The protocol defined by Algorithms~\ref{alg:(k-1)1}, \ref{alg:(k-1)2}, \ref{alg:(k-1)3}, and \ref{alg:(k-1)4} is a $(k-1)$-quiescent self-stabilizing solution to the $k$-gossip problem in arbitrary networks in the $(Synch, \textbf{CW}, *)$-model.
\end{theorem}

\noindent
\emph{Proof sketch.}
Let $p$ be the agent with the minimum $id$. From Lemma \ref{lem:minID}, eventually $p$ makes DFTs repeatedly over the whole network. Once $p$ completes the DFT, $MinID_v$ never becomes smaller than $p$ at any node $v$.

Now consider $p$'s DFT of the whole network that is initiated at a configuration satisfying $MinID_v \ge p$ at 
every node $v$. Then, $p$ repeatedly performs DFTs without waiting at any node, and $p$ completes each DFT in $2m$ rounds. This implies that timeout never occurs at any node starting from the second DFT. Any agent $q$ other than $p$ suspends its DFT when visiting any node $u$.  Agent $q$ can return to its suspended traverse only when timeout occurs at $u$.  However, since timeout never occurs at $u$, $q$ never returns to its suspended traverse and remains at $u$ forever.
\qed
\vspace{10pt}

To complete our results for the synchronous case, let us observe that in the $(Synch, \textbf{CW}, *)$ and $(Synch, \textbf{FW}, *)$ models, the quiescence number of the $k$-gossip problem among named agents is $k-1$ (by Theorems \ref{th:imp-k} and \ref{th:prt-(k-1)SC}). There remains the case of \textbf{NW} whiteboards, unfortunately the following theorem show that when the memory of agents is bounded (the bound may depend on the network size $n$), the $k$-gossip problem among named agents is not solvable.

\begin{theorem}
\label{th:qn-N}
The quiescence number of the $(*, \textbf{NW}, *)$-model is $-1$ for the $k$-gossip problem among named agents, when state space of each agent is bounded (but may depend on the network size $n$).
\end{theorem}

\noindent
\emph{Proof sketch.}
We prove the impossibility for synchronous ring networks. We assume for the purpose of contradiction that a 0-quiescent solution exists, and that each agent has at most $s$ states. Notice that $s$ is not necessarily a constant and may depend on the network size.

We consider system executions where each agent starts its execution from a predetermined state. Since no information can be stored in the whiteboards (model \textbf{NW}), the behavior of an agent depends solely on its own state and $id$ (the network being regular). 
When an agent executes an action, it changes its state then (potentially) moves (clockwise or counterclockwise).
Since each agent has at most $s$ states, it repeats a cyclic execution of at most length $s$ unless the agent meets another agent. Since only three kinds of moves are possible, there exists at most $3^{s+1}$ moving patterns in the cyclic behavior of length $s$ or less. Now we consider a sufficiently large domain of agent identifiers (\emph{e.g.} $k\times 3^{s+1}$). All possible agents are partitioned into at most $3^{s+1}$ groups depending on their moving patterns, and thus, some group contains $k$ or more agents. Now consider $k$ agents in the group of size $k$ or more, that are placed regularly in different nodes in the initial configuration of the nodes. Since agents in the group makes the same moving pattern in the cycle, the agents repeat the cyclic action without meeting each other in the synchronous execution. In the models with the whiteboards \textbf{NW}, the gossiping cannot be achieved without meetings of agents, which is a contradiction.
\qed
\vspace{10pt}

Note that the impossibility result holds even though the agents all start from a well known predefined initial state. Thus, if the initial location of agents is not controlled, even non-stabilizing solution are impossible to design.
For \emph{asynchronous} models, the remaining question is about the possibility of 0-quiescence.

\begin{theorem}
\label{th:qn-ACf}
The quiescence number of $(Asynch, \textbf{CW}, full)$ and $(Asynch, \textbf{NW}, full)$ model is $-1$ for the $k$-gossip problem among named agents.
\end{theorem}

\noindent
\emph{Proof sketch.}
We show that there exists no 0-quiescent self-stabilizing solution to the $k$-gossip problem in $(Asynch, \textbf{CW}, full)$-model. Let us assume for the purpose of contradiction that there exists a 0-quiescent self-stabilizing solution. All $k$ agents must keep moving in the 0-quiescent solution since 1-quiescence is impossible from Theorem \ref{th:imp-lA}.

Now consider a particular agent $p$. In the asynchronous system with \emph{full-duplex} links, there exists an execution such that $p$ never meets any other agent: before $p$ reaches a node $u$, all the agents staying at $u$ leave $u$. Notice that full-duplex links allow the agents to leave $u$ without meeting $p$: scheduling allows to have all agents exiting $u$ by the same link used by $p$ to arrive at $u$ to be moving concurrently with $p$. It follows that in the execution, agent $p$ cannot disseminate its own information (agents have to meet one another in \textbf{CW} model). Hence the result.
\qed
\vspace{10pt}

\begin{theorem}
\label{th:qn-AF}
The quiescence number of $(Asynch, \textbf{FW}, *)$-model is $0$ for the $k$-gossip problem among named agents.
\end{theorem}

\noindent
\emph{Proof sketch.}
Theorems \ref{th:imp-lA} shows that $1$-quiescence is impossible.  Thus, it is sufficient to present a $0$-quiescent self-stabilizing solution in $(Asynch, \textbf{FW}, *)$-model.

Consider the following protocol outline. Every agent repeatedly performs DFTs of the network. When an agent visits a node, it stores its gossip information in the whiteboard and collects the gossip information stored in the whiteboard.
After a DFT has been completed by every agent, all whiteboards contain the gossip information of all the agents, and every agent can obtain all the gossip information by performing an additional DFT.

The self-stabilizing DFT can be realized in the same way as the protocol presented in Theorem \ref{th:prt-(k-1)SC}: each agent simply behaves as the agent with the minimum $id$ of the protocol, yet does not need to wait at any node.  
\qed
\vspace{10pt}

\begin{theorem}
\label{th:qn-ACh}
The quiescence number of $(Asynch, \textbf{CW}, half)$-model is $0$ for the $k$-gossip problem among named agents.
\end{theorem}

\noindent
\emph{Proof sketch.}
Theorems \ref{th:imp-lA} shows that $1$-quiescence is impossible to attain.  Thus, it is sufficient to present a $0$-quiescent self-stabilizing solution in $(Asynch, \textbf{CW}, half)$-model.

Consider the following protocol outline. Every agent repeatedly performs DFTs of the network while recording at every traversed node the last targeted neighboring node. By the recorded information, other agents can trace a particular agent.
When an agent visits a node and finds a smaller $id$ than its own, it starts tracing the agent with the smaller $id$.
Eventually all the agent other than agent $p$ (that has minimal $id$) continue tracing $p$, that in turns perform a DFT forever. Since we assume the half-duplex edges, agents cannot miss one another on a link, and agents perform the same DFT and the agent with minimal $id$. Then a similar argument as in~\cite{BGT07c} implies that all agents other than $p$ meet $p$ infinitely often. Thus, by means of agent $p$, every agent can disseminate its gossip information to all other agents.

The self-stabilizing DFT can be realized in the same way as the protocol presented in Theorem \ref{th:prt-(k-1)SC}.
The only difference is in the way to detect the \emph{fake} $id$s. In the protocol of Theorem \ref{th:prt-(k-1)SC}, fake $id$s are detected by a timeout mechanism. Here, each agent records at each node the distance from the starting node in the depth-first-tree. In any trace labeled with a fake $id$, the tracing agent eventually detects contradiction in the distances and then decides that the traced $id$ is a fake one. Agent $p$ detecting a fake $id$ erases the false records on the path $p$ traced.
\qed
\vspace{10pt}

\section{Self-stabilizing $k$-gossip among anonymous agents}
\label{sec:anonymous}

Named agents being a stronger assumption than anonymous agents, all the impossibility results for named agents also hold for anonymous agents. In this section, we consider only the model variations that the impossibility results for named agents do not cover.

\begin{theorem}
\label{th:qn-SCanonym}
The quiescence number of $(*, \textbf{CW}, *)$-model is $-1$ for the $k$-gossip problem among anonymous agents.
\end{theorem}

\noindent
\emph{Proof sketch.}
Consider a synchronous ring network where all the whiteboards of nodes contain the same initial information. Assume that all the agents are in the same state in the initial configuration. In the synchronous system, all the agents move exactly the same and they never meet each other, and thus, the gossiping cannot be completed.
\qed
\vspace{10pt}

\begin{theorem}
\label{th:qn-AFanonym}
The quiescence number of $(Asynch, \textbf{FW}, *)$-model is $0$ for the $k$-gossip problem among anonymous agents.
\end{theorem}

\noindent
\emph{Proof sketch.}
From Theorem \ref{th:imp-lA}, it is sufficient to present a $0$-quiescent self-stabilizing solution to the gossip problem in $(Asynch, \textbf{FW}, *)$-model.

Since the whiteboards \textbf{FW} is available, the $k$-gossiping can be completed if every agent repeatedly traverses the network. However, an anonymous agent cannot record at a node that it has visited the node since its record cannot be distinguished from that of others. Thus, anonymous agents cannot execute the DFT like the ones in Theorem \ref{th:prt-(k-1)SC}.
Instead, each agent can traverse all the paths of a given length, say $\ell$, using the link labels (\emph{i.e.}, traverse all the paths in the lexicographic order of the label sequences). When completing the traverse of the paths of length $\ell$, the agent starts traversing the paths of length $\ell +1$. By repeating the traverses with incrementing the length, eventually the agent can traverse the whole network.
\qed
\vspace{10pt}

For the \emph{synchronous} anonymous agents, Theorem \ref{th:qn-AFanonym} guarantees that the quiescence number is at least $0$. On the other hand, the impossibility of $k$-quiescence for \emph{synchronous named} agents (Theorem \ref{th:imp-k}) leads to the following theorem.

\begin{theorem}
\label{th:qn-SFanonym}
The quiescence number of $(Synch, \textbf{FW}, *)$-model is not larger than $k-1$ and not smaller than $0$ for the $k$-gossip problem among anonymous agents.
\qed
\end{theorem}

\section{Conclusion}
\label{sec:conclusion}

This paper introduced the notion of quiescence for mobile agent protocols in a self-stabilizing setting. This notion complements the notion of silence~\cite{silent} used in ``classical'' self-stabilizing protocols. While $k$-quiescence of $k$-gossiping among named agents is easily attainable in \emph{non-stabilizing} solutions (assuming \textbf{FW} and 
\textbf{CW} whiteboards)~\cite{SIOKM08,SIOKM07}, this paper shows that \emph{self-stabilization} prevents $k$-quiescent solutions in any considered model, and even $0$-quiescent solutions in some particular models.   
Thus, our paper shed new light on the inherent difference between non-stabilizing and self-stabilizing solutions of
agent-based systems.

We would like to point out interesting open questions:
\begin{enumerate}
\item What is the exact quiescence number of the $(Synch, \textbf{FW}, *)$-model
for the $k$-gossip problem among \emph{anonymous} agents? (besides being not smaller than $0$ and not larger than $k-1$)
\item What is the connection between the quiescence number and the topology ?
\item Does there exist a non-trivial \emph{non-stabilizing} problem with quiescence number lower than $k$ ?
\end{enumerate}

\bibliographystyle{plain}
\bibliography{MobileRobot}

\begin{thebibliography}{10}

\bibitem{BHS01c}
J.~Beauquier, T.~Herault, and E.~Schiller.
\newblock Easy stabilization with an agent.
\newblock In {\em Proceedings of the 5th Workshop on Self-Stabilizing Systems
  (WSS)}, pages 35--51, 2001.

\bibitem{BGT07c}
L.~Blin, M.~Gradinariu Potop-Butucaru, and S.~Tixeuil.
\newblock On the self-stabilization of mobile robots in graphs.
\newblock In {\em Proceedings of OPODIS}, pages 301--314, 2007.

\bibitem{DDF07}
C.~Delporte-Gallet, S.~Devismes, and H.~Fauconnier.
\newblock Robust stabilizing leader election.
\newblock In {\em Proceedings of the 9th International Symposium on
  Stabilization, Safety, and Security of Distributed Systems (SSS)}, pages
  219--233, 2007.

\bibitem{D00b}
S.~Dolev.
\newblock {\em Self-stabilization}.
\newblock MIT Press, March 2000.

\bibitem{silent}
S.~Dolev, M.~G. Gouda, and M.~Schneider.
\newblock Memory requirements for silent stabilization.
\newblock {\em Acta Inf.}, 36(6):447--462, 1999.

\bibitem{DSW02c}
S.~Dolev, E.~Schiller, and J.~Welch.
\newblock Random walk for self-stabilizing group communication in ad-hoc
  networks.
\newblock pages 70--79, 2002.

\bibitem{G00c}
S.~Ghosh.
\newblock Agents, distributed algorithms, and stabilization.
\newblock pages 242--251, 2000.

\bibitem{HM01c}
T.~Herman and T.~Masuzawa.
\newblock Self-stabilizing agent traversal.
\newblock pages 152--166, 2001.

\bibitem{SIOKM08}
T.~Suzuki, T.~Izumi, F.~Ooshita, H.~Kakugawa, and T.~Masuzawa.
\newblock Move-optimal gossiping among mobile agents.
\newblock {\em Theoretical Computer Science (to appear)}.

\bibitem{SIOKM07}
T.~Suzuki, T.~Izumi, F.~Ooshita, H.~Kakugawa, and T.~Masuzawa.
\newblock Optimal moves for gossiping among mobile agents.
\newblock In {\em Proceedings of SIROCCO}, pages 151--165, 2007.

\end{thebibliography}

\end{document}